\documentclass{article} % For LaTeX2e
\usepackage{iclr2023_conference,times}

\usepackage{amsmath,amsfonts,bm}

\def\eqref#1{equation~\ref{#1}}
\def\1{\bm{1}}

\DeclareMathAlphabet{\mathsfit}{\encodingdefault}{\sfdefault}{m}{sl}
\SetMathAlphabet{\mathsfit}{bold}{\encodingdefault}{\sfdefault}{bx}{n}

\usepackage{hyperref}
\usepackage{url}
\usepackage{fontawesome}  % for GitHub icon

\usepackage{microtype}
\usepackage{graphicx}
\usepackage{booktabs} % for professional tables

\usepackage{algorithm}
\usepackage{algorithmic}

\usepackage{csquotes}
\usepackage{listings}
\usepackage{array}
\usepackage{tabularx}
\usepackage{multirow}
\usepackage{multicol}
\usepackage{hhline}
\usepackage{caption}
\usepackage{subcaption}
\usepackage{comment}
\usepackage{wrapfig}
\usepackage{tablefootnote}

\newcommand{\yy}{\mathbf{y}}

\newcommand{\xx}{\mathbf{x}}
\newcommand{\zz}{\mathbf{z}}

\newcommand{\githublink}{https://github.com/dvruette/figaro}
\newcommand{\soundcloudlink}{https://tinyurl.com/bdfds8f5}
\newcommand{\mixingsampleslink}{https://tinyurl.com/4tksvsub}
\newcommand{\moresampleslink}{https://tinyurl.com/2ps5n2nv}
\newcommand{\colablink}{https://tinyurl.com/28etxz27}

\title{FIGARO: Controllable Music Generation \\using Expert and Learned Features}

\author{Dimitri von~Rütte\thanks{Equal contribution},\, Luca Biggio\footnotemark[1],\, Yannic Kilcher, Thomas Hofmann  \\
Department of Computer Science, ETH Zürich\\
\texttt{dimitri.vonrutte@inf.ethz.ch} \\
}

\iclrfinalcopy % Uncomment for camera-ready version, but NOT for submission.
\begin{document}

\maketitle

\begin{abstract}
    Recent symbolic music generative models have achieved significant improvements in the quality of the generated samples. Nevertheless, it remains hard for users to control the output in such a way that it matches their expectation. To address this limitation, high-level, human-interpretable conditioning is essential. In this work, we release FIGARO, a Transformer-based conditional model trained to generate symbolic music based on a sequence of high-level control codes. To this end, we propose \emph{description-to-sequence} learning, which consists of automatically extracting fine-grained, human-interpretable features (the \emph{description}) and training a sequence-to-sequence model to reconstruct the original sequence given only the description as input. FIGARO achieves state-of-the-art performance in multi-track symbolic music generation both in terms of style transfer and sample quality. We show that performance can be further improved by combining human-interpretable with learned features. Our extensive experimental evaluation shows that FIGARO is able to generate samples that closely adhere to the content of the input descriptions, even when they deviate significantly from the training distribution.
\end{abstract}

\section{Introduction}
\label{introduction}

Music is a fascinating subject that surrounds us constantly, being a source of inspiration and canvas for imagination to many.
To some, creating music is a topic worthy of dedicating one's life to, which is a testament to the artistry and mastery involved.
While composition is an intricate form of art that requires a deep understanding of the human experience and domain knowledge, the idea of devising a systematic or algorithmic approach to music creation has been around for centuries \citep{nierhaus_historical_2009}.
With the advent of deep learning, automatic music generation has witnessed renewed interest \citep{hernandez-olivan_music_2021}.
% 1. Brief overview of previous/related work
%Especially the Transformer architecture \citep{vaswani_attention_2017}, popularized in the context of Natural Language processing \citep{brown2020language} and then successfully applied to several other Machine Learning domains \citep{dosovitskiy_image_2021,lample2019deep,pmlr-v139-biggio21a}, has proven to be a powerful tool for musical sequence modelling.
Especially the Transformer architecture \citep{vaswani_attention_2017}, which has seen applications to many Machine Learning domains \citep{brown2020language, dosovitskiy_image_2021,lample2019deep,pmlr-v139-biggio21a}, has proven to be a powerful tool for musical sequence modelling.
Initial breakthroughs by \citet{huang_music_2018} and \citet{payne_musenet_2019} applied language modelling techniques to symbolic music to achieve state-of-the-art music generation.
Though these models were capable of some form of conditional generation (e.g. melody or genre conditioning), other conditioning mechanisms and different types of control have since been proposed \citep{ens_mmm_2020, choi_encoding_2020, wu_musemorphose_2021}.
% e.g. bar- or track-level inpainting, musical style interpolation and fine-grained control

As deep generative models are improving and producing more and more realistic samples, it remains an area of active research how humans can interact with these models and steer them to generate a desirable result.
Recent efforts in text-to-image generation \citep{ramesh_hierarchical_2022, saharia_photorealistic_2022} have shown the potential in usability and artistic applications of human-interpretable controllable generative models.
Whereas text-based conditioning has yielded human-interpretable control for image generation, the same conditioning mechanisms are not easily applicable to music generation.
We aim to extend this kind of control to other domains, in this case to music generation.
As scale has proven to be key for achieving capable models, we cannot rely on scarce annotated data and instead propose a self-supervised objective, which we call \emph{description-to-sequence} learning.
We take inspiration from recent text-to-image approaches, but instead of a natural language description of the target, we automatically extract a sequence of high-level features (the \emph{description}). These can either be hand-crafted using domain knowledge or learned.
The description then serves as the input to a conditional model to reconstruct the original sequence.
To this end, we define a \emph{description function} which extracts said features from a given piece of music.
The choice of description function determines the characteristics of the resulting model and serves as an inductive bias, allowing us to emphasize desirable properties such as human-interpretability or fine-grained control over instruments and chord progression.
Note that the general nature of the proposed framework allows for applications to other domains despite our focus on symbolic music.

\begin{figure}
    \centering
    \includegraphics[width=\textwidth]{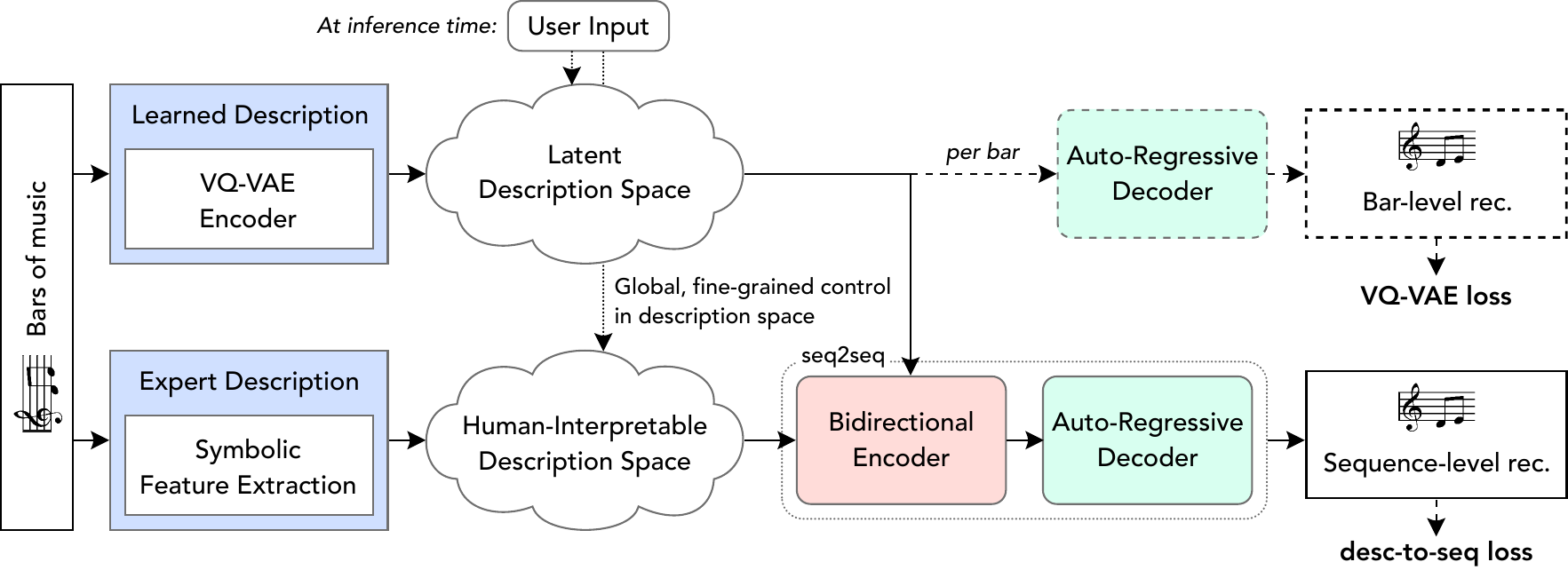}
    \caption{Overview of FIGARO.
        Dashed lines indicate components that are only used during training of the learned description.
        \vspace{-.4cm}
    }
    \label{fig:method_illustration}
\end{figure}

Our main contribution is FIGARO (FIne-grained music Generation via Attention-based, RObust control), a model trained on the proposed \emph{description-to-sequence} objective by combining two separate description functions:
1) The hand-crafted \emph{expert} description, which provides global context in the form of a high-level, human-interpretable sequence and 2) the \emph{learned} description, where we use representation learning to extract salient features from the source sequence.
The learned description is intended to amend the expert description with high-fidelity information in places where the latter might be incomplete, albeit at the cost of human-interpretability.
The model is trained on conditional generation, mapping descriptions to music.
An illustrated overview of the model is given in Figure~\ref{fig:method_illustration}.
At inference time, users may interact with the model in human-interpretable description space.
We provide a simple interface in the form of an online demo of our model.\footnote{Online demonstration of FIGARO is available on \href{\colablink}{Google Colab (\textit{\colablink})}. We recommend selecting a GPU environment for improved inference speed.}
We also release the source code and model weights for anyone to download and use freely.\footnote{Code and model weights are available through \href{\githublink}{GitHub (\textit{\githublink}).}}
Our secondary contribution is REMI+, an extension to the REMI input representation \citep{huang_pop_2020} which opens the way to multi-track, multi-time-signature music.

We evaluate FIGARO on its ability to adhere to the prescribed condition by comparing it to state-of-the-art methods for controllable symbolic music generation \citep{choi_encoding_2020, wu_musemorphose_2021}.
We demonstrate empirically that our technique outperforms the state-of-the-art in controllable generation and sample quality.
To evaluate sample quality, we employ subjective evaluation in the form of a listening study.
We further demonstrate that FIGARO is robust with respect to distributional shifts in description space and performs well on constructed samples outside the training distribution, indicating that the proposed objective is effective at learning generalized concepts about the data.

\section{Controllable Symbolic Music Generation}
In the context of generative modelling, controllability is an important issue, as such models only become useful if the user is able to steer the generation process in a desired direction.
This has recently been observed for text-to-image models and we intend to take a closer look at controllable music generation.
We identify two different levels of controllability: global and fine-grained control.
Global conditioning, where the generation is guided by a constant set of attributes that do not change during the generation process, is the most prevalent form of control.
Examples of global control include prompt-based conditioning \citep{payne_musenet_2019} or conditional decoding of latent representations \citep{brunner_midi-vae_2018}.
Fine-grained control is achieved when the generation process can be guided at any point in time, i.e. if the control attributes can be arbitrarily varied over time.
Consider controlling what instruments are playing in the generated sequence as an example of global control in contrast to controlling what instruments are playing \emph{at any point in time} as an example of fine-grained control.
Note that fine-grained control also implies global control, as global control can be achieved by fixing the control attributes to some constant value.
Fine-grained control is therefore a strictly more powerful property and seemingly harder to obtain, as is highlighted in Table~\ref{tab:method_comparison}.

\begin{table*}
    \centering
    \resizebox{\columnwidth}{!}{%
    \begin{tabular}{|l|l|cccc|}
        \hline
        Method & Input Rep. & Multi-Track & Multi-Sig. & Global Ctrl. & Fine-Grained Ctrl. \\
        \hline
        MIDI-VAE \citet{brunner_midi-vae_2018} & Pianoroll & \checkmark & \checkmark & \checkmark & - \\
        MuseNet \citep{payne_musenet_2019} & MIDI-like & \checkmark & \checkmark & \hspace{4pt}(\checkmark)\footnotemark\label{payne} & - \\
        MMM \citep{ens_mmm_2020} & MIDI-like & \checkmark & \checkmark & \checkmark & \hspace{4pt}(\checkmark)\footnotemark \\
        \citet{choi_encoding_2020} & MIDI-like & - & \checkmark & \checkmark & \hspace{4pt}(\checkmark)\footnotemark\label{choi} \\
        MuseMorphose \citep{wu_musemorphose_2021} & REMI & - & - & \checkmark & \checkmark \\
        \hhline{|=|=|====|}
        FIGARO (ours) & REMI+ & \checkmark & \checkmark & \checkmark & \checkmark \\
        \hline
    \end{tabular}
    }
    \caption{
        Generative capability and controllability comparison between different methods proposed in the literature.
        We compare our method to other state-of-the-art symbolic music generation methods on modelling capability (can the model generate multi-track/multi-time-signature music?) and controllability (can the generation be controlled on a global/fine-grained level?).
        \vspace{-.4cm}
    }
    \label{tab:method_comparison}
\end{table*}

\footnotetext[3]{{While it is possible to control the style (via artist tags) and instruments of the generated sequence, this information is \enquote{forgotten} by the model due to context scrolling once the generation advances beyond the initial prompt.}}
\footnotetext[4]{Fine-grained control is limited to changing the note density through bar-level inpainting.}

\footnotetext[5]{Fine-grained control is limited to optionally prescribing the melody of the generated sequence.}

\subsection{Related Work}
The capabilities of symbolic music generative models have been steadily improving with many recent contributions \citep{huang_music_2018, payne_musenet_2019, huang_pop_2020, hsiao_compound_2021, wu_power_2022}.
This line of work focuses on improving the quality of generated samples but does not contribute substantially toward controllability. An exception to that is MuseNet \citep{payne_musenet_2019}, which allows some control through prompt-based conditioning with control tokens.
Even still, prompt-based control is very limited, as control tokens are \enquote{forgotten} by the model once the generation advances beyond the initial context size.
\\
Another line of work focuses on finding ways of controlling the generation process.
VAE-based methods often achieve global control through latent conditioning vectors, enabling genre transfer \citep{brunner_midi-vae_2018} capabilities or control over arousal \citep{tan_music_2020}.
Transformer auto-encoders have been used for melody conditioning \citep{huang_music_2018, choi_encoding_2020} and encoding of musical style \citet{choi_encoding_2020}.
\citet{ens_mmm_2020} present MMM which is capable of bar-level and track-level symbolic music inpainting.
Finally, \citet{wu_musemorphose_2021} introduce MuseMorphose for fine-grained attribute conditioning and latent space style editing.
All of these approaches have some limitations that are highlighted in Table~\ref{tab:method_comparison}.
Recent work tackling the problem of controllable generation is often either limited to single-track, single-time-signature music or to global control.
\\
Fine-grained control has been a topic of interest in the recent literature \citep{choi_encoding_2020, wu_musemorphose_2021, di_video_2021, ferreira_learning_2021} and is an essential property when considering user-directed applications.
In essence, fine-grained control is necessary to allow control over salient features in the generation, as saliency in music at least partly lies in how it changes over time.
In addition, salient features may be impossible or prohibitively expensive to quantify directly \citep{choi_encoding_2020, ferreira_learning_2021}, emphasizing the need for un- or self-supervised fine-grained control. 
Our proposed method provides both global and fine-grained control through description-to-sequence learning.

\section{Description-to-Sequence Learning}
\label{sec:description_to_sequence}

Description-to-sequence learning is a self-supervised objective based on the reconstruction of a sequence given automatically extracted features of that same sequence (the \emph{description}) as input.
Conceptually, an information bottleneck is applied by mapping a sample to the description space and the model is trained to learn a probabilistic map from description space to sample space.

\begin{wrapfigure}{r}{0.5\textwidth}
    \centering
    \vspace{-.55cm}
    \includegraphics[width=0.49\columnwidth]{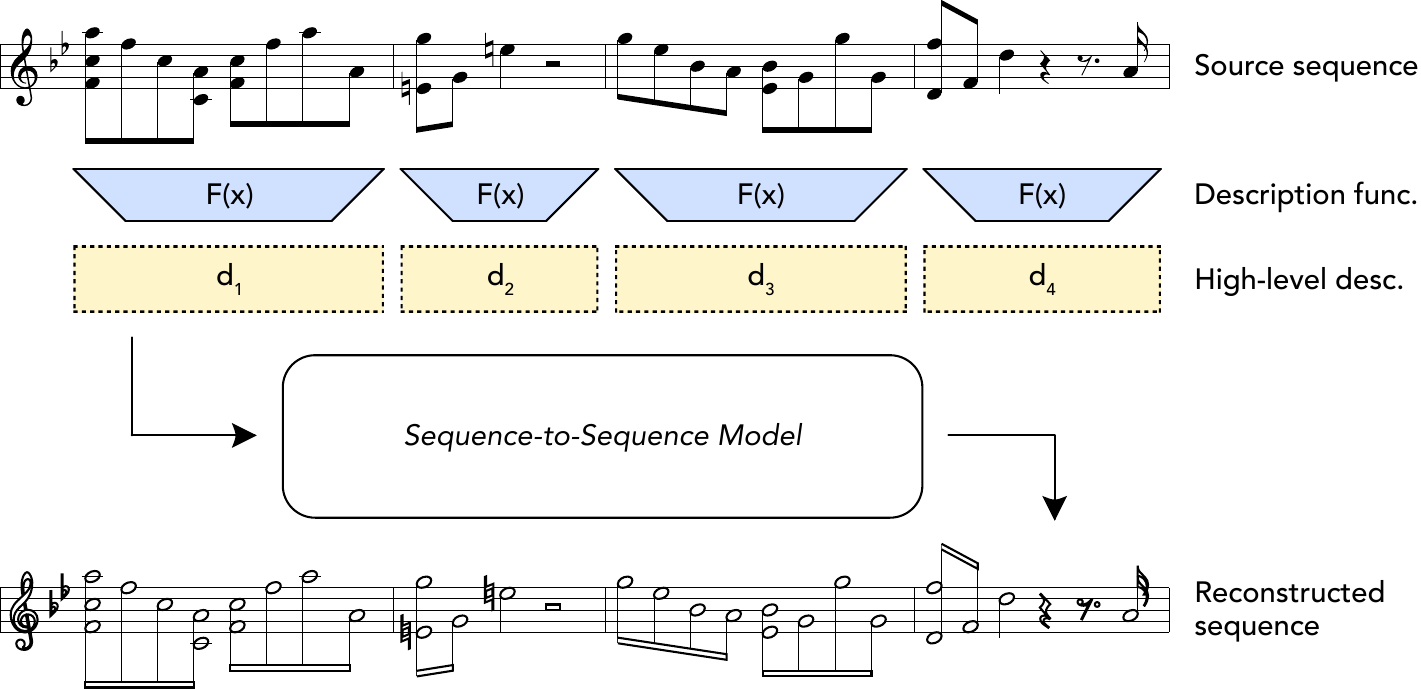}
    \caption{Schematic overview of description-to-sequence learning in the context of music.\vspace{-.15cm}}
\centering
\tiny
\label{fig:desc_to_seq_illustration}
\end{wrapfigure}

\textbf{Description-to-Sequence Objective.}
Let $V_\mathrm{seq}$ denote a sequence vocabulary and $V_\mathrm{desc}$ a description vocabulary.
Let $\xx \in V_\mathrm{seq}^*$ be an input sequence and $F: V_\mathrm{seq}^* \to V_\mathrm{desc}^*$ be a partition-wise feature extraction function that extracts descriptive features given a partition $x_1, \dots, x_n$ of $\xx$ such that $\xx = \mathrm{concat}(x_1, \dots, x_n)$.
We then define the description of $\xx$ to be:
\begin{equation}
    \mathbf{d} = \mathrm{concat}(F(x_1), \dots, F(x_n))
\end{equation}
For simplicity, we write $\mathbf{d} = F(\xx)$ in the following.
The description-to-sequence objective is then given by the cross-entropy reconstruction loss of $\xx$ given $F(\xx)$:
\begin{equation}
    \label{eq:rec_loss}
    \mathcal{L}_\mathrm{rec}(\phi) = \mathbb{E}_{\xx \sim \mathcal{X}}[-\log p_\phi(\xx | F(\xx))]
\end{equation}
Here $\mathcal{X}$ is the true distribution of sequences and $\phi$ denotes the parameters of a sequence-to-sequence model. An illustrated overview of this objective is given in Figure~\ref{fig:desc_to_seq_illustration}.\\
Note that some level of fine-grained control is guaranteed by the way we partition $\xx$ prior to feature extraction, as the length of partitions puts an upper bound on the receptive field of the description function.

\vspace{-.1cm}
\subsection{Expert Description}
\vspace{-.15cm}
\begin{figure}
    \centering
    \begin{tabular}{l}
    \scriptsize
\begin{lstlisting}
<bos>
  Bar_1 TimeSignature_4/4 NoteDensity_3 MeanPitch_14 MeanVelocity_19 MeanDuration_32
    Instrument_Drums Instrument_Piano Instrument_E-Piano Instrument_SlapBass
    Chord_E:maj Chord_F#:min7
  Bar_2 TimeSignature_4/4 NoteDensity_3 MeanPitch_15 MeanVelocity_18 MeanDuration_27
    Instrument_Drums Instrument_Piano Instrument_SlapBass Instrument_Guitar
    Chord_C#:min Chord_E:maj Chord_A:maj7
  Bar_3 ...
<eos>
\end{lstlisting}
    \vspace{-.1cm}
    \end{tabular}
    \caption{
        Example of an expert description.
        The description contains information about time signature, note density, pitch, velocity and duration as well as which instruments and chords are played throughout each bar.
        \vspace{-.4cm}
    }
    \label{fig:description_example}
\end{figure} 
We aim for a human-interpretable, fine-grained conditioning mechanism by using domain knowledge to automatically extract high-level features that are easy to understand for human experts and relevant from a compositional standpoint.
We identify time signature, chords and instruments as such features and also consider note density, mean pitch, mean velocity and mean duration as auxiliary features quantifying musical style \citep{choi_encoding_2020}.
All of these features are computed bar-per-bar and quantized according to a special-purpose vocabulary. Finally, all resulting tokens are concatenated into a sequence, which yields the expert description function $F_\mathrm{expert}(\xx)$, of which an example is given in Figure~\ref{fig:description_example}.
More details on the feature extraction algorithm and quantization can be found in Appendix~\ref{description_details}.

\vspace{-.1cm}
\subsection{Learned Description}
\vspace{-.15cm}
While the expert description serves as a human-interpretable inductive bias, it is only able to capture fairly low-fidelity features by design and suffers from non-injectivity (i.e. there can be many sequences that map to the same description).
In an attempt to alleviate some of these limitations, we use representation learning to extract features that maximize the information content about the underlying sequence.
We choose the VQ-VAE framework as the basis for our learned description function as it produces discrete codes that allow for learning a prior sequence model and has been shown to be effective for various tasks such music generation in particular \citep{dhariwal_jukebox_2020}.
The latent representations given by this model is used as the output of the learned description function $F_\mathrm{learned}(\xx)$.
This way, we add high-fidelity information and reduce non-injectivity, albeit at the cost of human interpretability.\\
We later show that combining both descriptions yields better performance than using either one on its own (Section~\ref{sec:conditional_generation}) and that the resulting model is fairly robust (Section~\ref{sec:zero_shot}).

\vspace{-.1cm}
\section{Experimental Setup}
\vspace{-.15cm}

\textbf{REMI+ Input Representation.}
The Revamped MIDI (REMI) representation \citep{huang_pop_2020} is a beat-based input representation, which has recently been shown to give better sample quality than traditional time-based representation.
REMI represents notes with four consecutive tokens encoding note position, pitch, velocity and duration and also includes chord and tempo events.

In order to make it applicable to any general MIDI piece, we extend REMI by adding instrument and time-signature information, yielding REMI+.
We add tokens encoding the time signature of a given bar at the beginning of each bar and instrument tokens at the beginning of each note, thus making it suitable for multi-time-signature, multi-track modelling tasks.
We also increase the temporal resolution of note positions from 4 to 12 sub-beats per quarter note.
This allows for precise quantization of triplets in addition to sixteenths, which was previously not possible.
For exact details on quantization and tokenization, we refer to Appendix~\ref{app:remi_plus}.
A limitation left unaddressed by REMI+ is the inability to encode expressive performances, which requires frequent shifts in tempo or small time offsets.
Time-based representations are still preferred for that setting.

In our experiments, we tokenize MIDI files using the REMI+ input representation and train FIGARO on the proposed objective using the combination of the proposed \emph{expert} and \emph{latent} description functions.

\textbf{Description-to-Sequence Model.}
We use the original Transformer auto-encoder \citep{vaswani_attention_2017} architecture with 4 encoder and 6 decoder layers as the sequence-to-sequence backbone.
Three separate models are trained using the proposed description-to-sequence objective (Equation~\ref{eq:rec_loss}) using either $F_\mathrm{expert}(\xx)$, $F_\mathrm{learned}(\xx)$ or a combination of both descriptions.
Combining descriptions is done by simply summing their respective learned token embeddings.
The model architecture is kept deliberately simple, so as to be able to accurately assess the effectiveness of the various description functions.
Further details and hyperparameters are given in Appendix~\ref{app:implementation_details}.

\textbf{VQ-VAE Model.}
For our bar-level VQ-VAE model, we use the same Transformer auto-encoder backbone but with an additional quantization bottleneck consisting of 16 latent codes with a shared codebook of size 2048 between the encoder and decoder.
The model is trained on reconstruction for individual bars.
Note that training FIGARO is a two-stage process:
In the first step, we train the VQ-VAE for feature extraction, which is later used to compute $F_\mathrm{learned}(\xx)$.
The VQ-VAE model is frozen during training of the description-to-sequence model.

\textbf{Training Setup.}
We use the LakhMIDI dataset \citep{raffel_learning-based_2016} as training data in all of our experiments, which to the best of our knowledge is the largest publicly available symbolic music dataset.
We use a 80\%-10\%-10\% training-validation-test split.
Each model is trained for 24 hours on 4 Nvidia GTX 2080 Ti GPUs.
For evaluation, we generate samples conditioned on descriptions sampled from the test set, generating 32 bars for each sample.
For training details, we refer to Appendix~\ref{app:implementation_details}.
%We additionally release the source code and trained model weights that were used in our experiments.\footnotemark

%\footnotetext{Source code and model weights are available on \href{\githublink}{GitHub}.}

\vspace{-.1cm}
\section{Performance Evaluation}
\label{sec:evaluation_metrics}

\vspace{-.15cm}
\subsection{Fluency}
\vspace{-.15cm}
We use perplexity (PPL) as a way to measure fluency and to compare the likelihood of different models in addition to task-specific metrics.
The perplexity measures the likelihood of sequences while normalizing over the sequence length, which makes it better suited to comparing sequences of different lengths than the negative log-likelihood.

\vspace{-.1cm}
\subsection{Description Fidelity}
\vspace{-.15cm}
We also quantitatively evaluate the fidelity of generated sequences to the given condition.
Let $\xx$ denote a test sample and $F(\xx)$ its description.
Then we generate $\hat{\xx}$ by sampling the model conditioned on $F(\xx)$ and examine $\xx$ and $\hat{\xx}$ for similarity.
Metrics are computed as an empirical estimate over the test distribution.
More details and exact formulas are given in Appendix~\ref{app:evaluation_metrics}.

\textbf{Accuracy.}
We compute accuracy metrics for categorical values, namely for instruments, chords and time signature.
Instruments and chord are multi-label features for which we compute the mean $F_1$ score.
As there is only one time signature per bar, we compute the standard accuracy score.

\textbf{Macro Overlapping Area.}
Previous work has used the overlapping area (OA) metric to quantify similarity between two musical sequences for a given feature \citep{choi_encoding_2020, wu_musemorphose_2021}.
However, we find that the standard OA metric fails to take the order of the sequences into account, as feature histograms are computed over the entire sequence.
For example, a sequence has maximal overlapping area with itself in reverse order, even though the resulting sequences will sound completely different.
To alleviate this limitation, we propose the macro overlapping area (MOA), which partition-wise computes the overlap in the distributions of a given feature, taking sequential order into account.
We use the MOA metric to compute similarity in pitch, velocity and duration between ground truth and reconstruction.
The exact definition of our MOA metric is given in Appendix~\ref{app:evaluation_metrics}.

\textbf{Normalized Root-Mean-Square Error.}
We compute the normalized RMSE (NRMSE) for bar-wise note density.
This helps compare similarity accross different feature magnitudes.

\textbf{Cosine Similarity.}
We also evaluate chroma and grooving similarity as a way to quantify similarity in sound and rhythm as proposed by \citet{wu_musemorphose_2021}.
We compute bar-wise cosine similarity for the chroma vectors \citep{fujishima_real-time_1999} and grooving vectors \citep{dixon_towards_2004}.

\vspace{-.1cm}
\section{Experiments}
\vspace{-.15cm}
Recent work on controllable generation has largely focused on single-track symbolic music and to the best of our knowledge, there are no competitive baselines for multi-track music generation with fine-grained control.
To make due, we train two state-of-the-art methods for single-track controllable generation \citep{choi_encoding_2020, wu_musemorphose_2021} on the REMI+ input representation, which allows us to apply them to a multi-track setting.
We also train an unconditional baseline based on Music Transformer \citep{huang_music_2018}, which acts as a sanity check:
It has no additional information about the target sequence and essentially performs \enquote{random guessing}, i.e. it uniformly samples a sequence from the training distribution.
If any conditional model performs worse than unconditional on similarity metrics, it is an indication for mode collapse \citep{dieng_avoiding_2018}.
Indeed, although the training of MuseMorphose \citep{wu_musemorphose_2021} initially appeared to be successful, quantitative analysis revealed that the model had suffered mode collapse, as evidenced by the small chord and instrument entropy in Table~\ref{tab:diversity_metrics} and the worse-than-unconditional performance on some of the style transfer metrics in Table~\ref{tab:cond_generation_results}.
As this model was originally designed for modelling single-track piano music and the LakhMIDI dataset constitutes a quite drastic domain shift, we instead focus on comparison to \citet{choi_encoding_2020}.
Details on baseline training are given in Appendix~\ref{app:benchmark_models}.

\vspace{-.1cm}
\subsection{Human Input}
\vspace{-.15cm}
The question of how well FIGARO can handle human-generated descriptions is both important and tricky to evaluate quantitatively, as both $F_\mathrm{expert}$ and especially $F_\mathrm{learned}$ require time and domain knowledge to create (albeit substantially less than composing a piece of music from scratch).
We therefore demonstrate the generative capabilities of our model in an interactive demo\footnote{Online demonstration of FIGARO on \href{\colablink}{Google Colab (\textit{\colablink})}} based on $F_\mathrm{expert}$.
We provide four hand-crafted examples, namely a simple four-bar piano progression, an 8-bar description containing uncommon instruments, a 9-bar description based on a well-known Jazz standard and a longer 20-bar description demonstrating control over multiple features in parallel with fine-grained instructions over instruments, chords, note density and mean pitch.
These example are meant to show several attributes of FIGARO, namely its level of controllability and its ability to generate meaningful outputs even when input descriptions are human-made and present features that rarely appear in the training set.
The reader is invited to generate samples based on these descriptions or to come up with their own creations.
Demonstrating the use of $F_\mathrm{learned}$ on human input is less straight-forward as in the absence of a prior model, latent codes need to be extracted from an existing piece of music.
We approximate this and the previous setting by using pieces of music from the test set to extract descriptions and perform quantitative evaluation on style transfer and controllable generation in Section~\ref{sec:conditional_generation} and~\ref{sec:zero_shot} respectively.

\vspace{-.1cm}
\subsection{Style Transfer}
\label{sec:conditional_generation}
\vspace{-.15cm}
In this experiment, we extract conditioning sequences by feeding samples drawn from the test set uniformly at random through neural bottlenecks in the case of MuseMorphose, \citet{choi_encoding_2020} and $F_\mathrm{learned}$ or a symbolic bottleneck in the case of $F_\mathrm{expert}$.
New samples are then generated based on the condition and closeness in style to the original sequence is evaluated using the metrics from Section~\ref{sec:evaluation_metrics}.
Note that many of these metrics are directly present in $F_\mathrm{expert}$ and FIGARO (expert) expectedly does well in that regard.
The fact that the results obtained with the expert description adhere to the content of the description itself is a highly desirable feature for controllable music generation.
However, it also performs well on chroma and grooving similarity, both of which are not directly present.
\begin{wraptable}{r}{0.5\textwidth}
    \centering
    \begin{tabular}{|l|c|c|}
        \hline
        Model & $\Delta H_\mathrm{inst}$ & $\Delta H_\mathrm{chord}$ \\
        \hline
        Ground truth & $(3.763)$ & $(4.077)$ \\
        \hline
        Unconditional & $-0.521$ & $-0.328$ \\
        \citet{choi_encoding_2020} & $-0.116$ & $-0.128$ \\
        \citet{wu_musemorphose_2021} & $-1.954$ & $-1.657$ \\
        \hline
        FIGARO (expert) & $-0.031$ & $-0.028$ \\
        FIGARO (learned) & \hspace{8pt}$0.062$ & \hspace{8pt}$0.008$ \\
        FIGARO & $-0.102$ & $-0.006$ \\
        \hline
    \end{tabular}
    \caption{Difference in instrument entropy $H_\mathrm{inst}$ and chord entropy $H_\mathrm{chord}$ between ground truth and modelled distributions. Model entropies are empirical estimates over samples from the style transfer task. Ground truth values are absolute, deltas are relative to ground truth.\vspace{-.2cm}}
    \label{tab:diversity_metrics}
\end{wraptable}
This shows that the expert description can act as a strong inductive bias and helps guide the generation process even beyond what information is directly represented.
By using this description, the degree of control exerted by the user is enhanced and the results match expectations.
We also observe that adding learned features further improves performance across the board, with FIGARO beating FIGARO (expert) in every category.
The success of this hybrid approach means that we can preserve the interpretability and inductive bias, yet increase the fidelity of the generated music by exploiting black-box AI.
Our expert description and hybrid models significantly outperform all baselines, and the learned description model outperforms \citet{choi_encoding_2020} on most metrics by a slight margin.
The difference in performance is explained by the conditioning used for \citet{choi_encoding_2020}, where the conditioning vector is temporally aggregated over the entire sample with any style progression throughout the sequence being lost.
We provide a non-cherrypicked collection of samples and encourage the reader to get an impression of the quality and diversity of the music by listening to some of them on \href{\soundcloudlink}{SoundCloud}\footnote{\href{\soundcloudlink}{\soundcloudlink}}.
The full list of results is provided in Table~\ref{tab:cond_generation_results}.
{
\setlength{\tabcolsep}{0.5em}
\begin{table*}[t]
    \begin{subtable}[h]{\textwidth}
        \centering
        \resizebox{\columnwidth}{!}{%
        \begin{tabular}{|m{3cm}|c|ccc|cccccc|}
            \hline
            \multirow{2}{3em}{Model} & Fluency & \multicolumn{3}{c|}{Accuracy} & \multicolumn{6}{c|}{Fidelity} \\
            & PPL\;$\downarrow$ & I\;$\uparrow$ & C\;$\uparrow$ & TS\;$\uparrow$ & ND\;$\downarrow$ & P\;$\uparrow$ & V\;$\uparrow$ & D\;$\uparrow$ & $s_c\uparrow$ & $s_g\uparrow$ \\
            \hline
            Unconditional & 1.988 & .191 & .048 & .751 & 2.192 & .563 & .153 & .312 & .306 & .510 \\
            \citet{choi_encoding_2020} & 2.049 & .658 & .184 & .908 & 1.679 & .646 & .574 & .484 & .514 & .688 \\
            \citet{wu_musemorphose_2021} & 2.094 & .179 & .050 & \textbf{1.00} & 0.873 & .492 & .050 & .207 & .312 & .529 \\
            \hline
            FIGARO (expert) & 1.913 & \textbf{.957} & .561 & \textbf{.996} & 0.319 & .759 & .658 & .514 & .712 & .637 \\
            FIGARO (learned) & 1.973 & .594 & .195 & .969 & 0.738 & .701 & .653 & .546 & .544 & .697 \\
            FIGARO & \textbf{1.705} & \textbf{.960} & \textbf{.593} & \textbf{.997} & \textbf{0.238} & \textbf{.827} & \textbf{.735} & \textbf{.748} & \textbf{.790} & \textbf{.853} \\
            \hline
        \end{tabular}
        }
        \caption{Perplexity and similarity metrics for style transfer on LakhMIDI. Best values are highlighted.\vspace{0.1cm}}
        \label{tab:cond_generation_results}
    \end{subtable}
    \begin{subtable}[h]{\textwidth}
        \centering
        \begin{tabular}{|m{3cm}|c|ccc|cccccc|}
            \hline
            \multirow{2}{4em}{Model} & Fluency & \multicolumn{3}{c|}{Accuracy} & \multicolumn{6}{c|}{Fidelity} \\
            & PPL\;$\downarrow$ & I\;$\uparrow$ & C\;$\uparrow$ & TS\;$\uparrow$ & ND\;$\downarrow$ & P\;$\uparrow$ & V\;$\uparrow$ & D\;$\uparrow$ & $s_c\uparrow$ & $s_g\uparrow$ \\
            \hline
            \citet{choi_encoding_2020} & 2.213 & .441 & .129 & .808 & 1.407 & .603 & .396 & .448 & .437 & .643 \\
            \hline
            FIGARO (expert) & 1.824 & \textbf{.944} & \textbf{.524} & \textbf{.992} & 0.384 & .741 & .559 & .497 & .705 & .575 \\
            FIGARO (learned) & 2.186 & .381 & .128 & .829 & 0.831 & .649 & .424 & .478 & .446 & .614 \\
            FIGARO & \textbf{1.782} & .917 & .514 & \textbf{.988} & \textbf{0.335} & \textbf{.807} & \textbf{.702} & \textbf{.694} & \textbf{.748} & \textbf{.744} \\
            \hline
        \end{tabular}
        \caption{Description splicing perplexity and similarity metrics. Best values are highlighted.\vspace{0.1cm}}
        \label{tab:desc_transfer_results}
    \end{subtable}
    \begin{subtable}[h]{\textwidth}
        \centering
        \begin{tabular}{|m{3cm}|c|ccc|cccccc|}
            \hline
            \multirow{2}{4em}{Model} & Fluency & \multicolumn{3}{c|}{Accuracy} & \multicolumn{6}{c|}{Fidelity} \\
            & PPL\;$\downarrow$ & I\;$\uparrow$ & C\;$\uparrow$ & TS\;$\uparrow$ & ND\;$\downarrow$ & P\;$\uparrow$ & V\;$\uparrow$ & D\;$\uparrow$ & $s_c\uparrow$ & $s_g\uparrow$ \\
            \hline
            FIGARO (expert) & \textbf{1.894} & \textbf{.955} & \textbf{.553} & \textbf{.996} & \textbf{0.360} & \textbf{.700} & \textbf{.646} & \textbf{.434} & \textbf{.710} & \textbf{.639} \\
            \hline
            - w/o instruments & 1.980 & \emph{.373} & \textbf{.568} & \textbf{1.00} & 0.424 & .674 & .586 & \textbf{.436} & .687 & \textbf{.625} \\
            - w/o chords & \emph{2.023} & .895 & \emph{.100} & \textbf{.995} & 0.564 & .672 & .603 & .413 & \emph{.294} & .615 \\
            - w/o meta info. & 1.966 & .908 & .536 & \emph{.795} & \emph{0.878} & \emph{.574} & \emph{.205} & \emph{.334} & .636 & \emph{.584} \\
            \hline
        \end{tabular}
        \caption{Ablation study perplexity and similarity metrics. Best/worst values are highlighted/emphasized.\vspace{-0.15cm}}
        \label{tab:ablation_study_results}
    \end{subtable}
    \caption{
        We compare our models to the unconditional baseline based on \citet{huang_music_2018}, \citep{choi_encoding_2020} and MuseMorphose \citep{wu_musemorphose_2021} on perplexity (PPL) and similarity metrics.
        Similarity metrics include instrument $F_1$-score (I), chord $F_1$-score (C) and time signature accuracy (TS) as well as note density NRMSE (ND), pitch MOA (P), velocity MOA (V), duration MOA (D), chroma similarity $s_c$ and grooving similarity $s_g$.
        \vspace{-0.5cm}
    }
\end{table*}
}

\vspace{-.2cm}
\subsection{Controllable Generation}
\label{sec:zero_shot}
\vspace{-.15cm}
To evaluate the robustness of our model with respect to out-of-distribution conditions, we combine two sequences $\xx^{(1)}$ and $\xx^{(2)}$ (either splicing or mixing the descriptions) and generate samples conditioned on the resulting descriptions.
This way, we create examples that have unusual transitions (in the case of splicing) or conflicting information (in the case of mixing) and are not part of the training distribution in general.

\textbf{Description splicing.}
For this experiment, we take 16 bars of each sequence and concatenate them into a novel description $\Tilde{\xx} = b_{1}^{(1)} \parallel \dots \parallel b_{16}^{(1)} \parallel b_{17}^{(2)} \parallel \dots \parallel b_{32}^{(2)}$.
We then input the description $F(\Tilde{\xx})$ to the model and sample the output distribution to generate a \enquote{medley} of the input sequences.
Importantly, none of the models are finetuned for this specific task.
We see a drop in performance in all evaluation metrics for every model compared to the style transfer task, which is expected due to distributional shifts in the data.
However, FIGARO and FIGARO (expert) drop significantly less than FIGARO (learned) and \citet{choi_encoding_2020}, showing that the expert description provides a strong inductive bias and is robust with respect to distribution shifts.
We sometimes even observe FIGARO turning the hard cut-off points from bar-splicing into smooth transitions, which can be attributed to its capacity to pay attention to future bars and anticipate significant changes in style.
The complete list of results is available in Table~\ref{tab:desc_transfer_results}.
We omit \citet{wu_musemorphose_2021} from these experiments as we do not expect competitive performance due to mode collapse.

\textbf{Description mixing.}
Description mixing is done by combining $F_\mathrm{expert}(\xx^{(1)})$ with $F_\mathrm{learned}(\xx^{(2)})$ and thus is a scenario that is only applicable to FIGARO, the only model that takes both description types as input.
Note that mixing descriptions in this way potentially provides conflicting information to the model as the two sequences might not agree with each other on certain features which might decrease sample quality or lead to a collapse of the model's performance in the worst case.
Conflicting information also makes quantitative evaluation more difficult, which is why we focus on qualitative evaluation.
We find that mixing descriptions in this way is not detrimental to the sample quality and that the model relies on different descriptions for different attributes of the generated sample.
In general terms, the model seems to rely on the expert description for attributes including instruments, harmonics and time signature (one might call these \enquote{low-fidelity} attributes) and on the learned description for attributes including texture and rhythmic intensity (\enquote{high-fidelity} attributes).
This is not surprising as the expert description is designed to make this kind of low-fidelity information readily available and the learned description is intended to \enquote{fill in the gaps} of the expert description.
We invite the reader to examine how the models combines two pieces of music into one by listening to some of the samples on \href{\mixingsampleslink}{SoundCloud}.
We also provide evaluation metrics with respect to the two source samples in Appendix~\ref{app:description_mixing}.

\vspace{-.125cm}
\subsection{Ablation Study}
\vspace{-.175cm}

To evaluate which parts of the expert description are essential, we group it into three components: instruments, chords and meta-information.
Instruments and chords include all tokens with information about instruments and chords respectively while all other tokens (time signature, note density and mean pitch, velocity and duration) are classified as meta-tokens.
We train separate models with one part of the description removed and compare the performance to FIGARO (expert), which receives the full expert description as input.\\
As one would expect, removing each component reduces the performance significantly in the respective metrics, indicating that each component carries useful information not entirely inferable through the remaining components.
Interestingly, our experiments show that removing any component slightly decreases the over-all performance even in metrics that we would not necessarily expect to be affected.
Removing instrument information, for example, increases the error for note density, pitch, velocity and duration, indicating that the instruments also carry mutual information about those features.
This seems plausible considering the fact that different styles (or genres) of music usually identifies a set of instruments that is common for said style.
Similar arguments can be made for the other two components.
The full list of results is available in Table~\ref{tab:ablation_study_results}.

\vspace{-.125cm}
\subsection{Subjective Evaluation}
\vspace{-.175cm}
\label{sec:subjective_evaluation}

\begin{wrapfigure}{r}{0.5\textwidth}
    \vspace{-0.1cm}
    \includegraphics[width=0.5\textwidth]{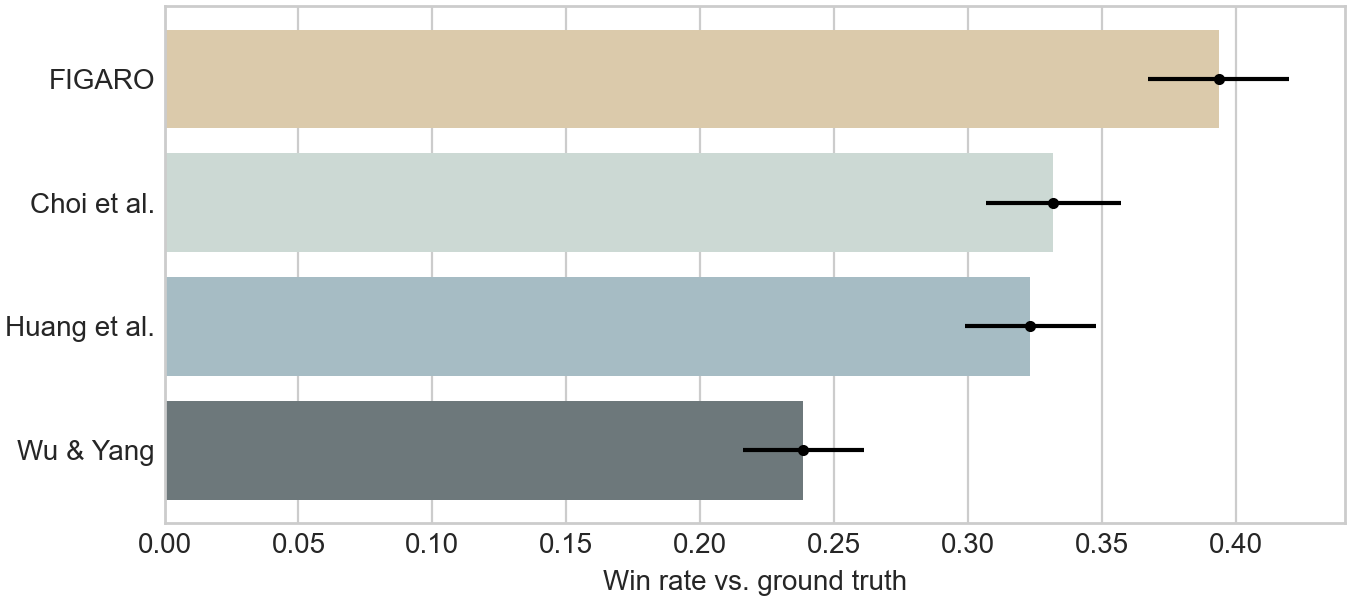}
    \caption{
        Win rates of generated samples against real samples. 
        We compare FIGARO, \citet{huang_music_2018}, \citet{choi_encoding_2020} and \citet{wu_musemorphose_2021}. 
        Real samples are from the test set. 
    }
    \label{fig:subjective_evaluation}
\centering
\end{wrapfigure}

Finally, we evaluate the subjective quality of generated samples through a listening study, comparing our best model to the baselines.
In this experiment we are not interested in how closely the generated samples follows the prescribed condition, but instead try to quantify the perceived quality of samples by recording pairwise preferences.
To this end, we have conducted a survey where participants were asked to indicate their preference between 20 second excerpts of two samples chosen uniformly at random.\footnote{Samples used in the study are available for download: \href{\moresampleslink}{\textit{\moresampleslink}}}
In total, we gathered 7569 comparisons by 691 unique participants, whose  expertise ranged everything between non-musicians and professional musicians, although no prior musical knowledge was required.
In two types of questions, participants had to choose 1) between a real sample and a generated sample or 2) between two generated samples. Generated samples are taken from the style transfer task.
Question type 1) ranks the different methods on how good generated samples are compared to real, human-composed music.
In this respect, FIGARO beats all other baseline with a win rate of $39.3\%$ compared to the next best model by \citet{choi_encoding_2020}, which has a win rate of $33.2\%$.
As evidenced by the fact that the unconditional baseline \citep{huang_music_2018} and \citet{choi_encoding_2020} are very close in terms of win rates, sample quality and reconstruction accuracy are not necessarily correlated, highlighting the need for a dedicated evaluation of sample quality.
Win rates of the different models are displayed in Figure~\ref{fig:subjective_evaluation} with $90\%$ confidence intervals obtained 
through normal approximation.
While \citet{choi_encoding_2020} is able to adhere to the prescribed condition as shown in Section~\ref{sec:conditional_generation}, the quality of generated samples is approximately on par with the unconditional model.
FIGARO on the other hand is able to surpass the unconditional baseline in sample quality, while also providing controllable generation capabilities.
Question type 2) is used to construct a pairwise ranking of models by applying the Wilcoxon signed-rank test on the study results. 
In this ranking, our model beats each of the baselines with a $p$-value of $< 10^{-7}$.
The complete ranking and test results can be found in Appendix~\ref{app:subjective_evaluation}.

\vspace{-.1cm}
\section{Discussion}
\vspace{-.15cm}
\textbf{Future Work.}
While our work represents a step towards high-quality symbolic music generation with human-interpretable control, we recognize three main avenues for future work.
Firstly, the proposed model is strictly speaking not a generative model as creative input in the form of a description is still required.
Training a prior model to generate such descriptions unconditionally could allow for faster generation with less human input required while still retaining the ability to influence the generation process in description space.
Secondly, adding new description functions or extending the proposed ones  could lead to further improved controllability and sample quality.
Melody conditioning stands out as a useful attribute still lacking control in FIGARO.
Lastly, applying the description-to-sequence objective to other domains such as natural language could potentially enable human-interpretable control in a way that is not currently possible. 
For example, the long-range conditioning ability could be interesting in the setting of story generation.

\textbf{Conclusion.}
% 1. our method works surprisingly well in the context of symbolic music generation. 
We present the self-supervised description-to-sequence objective and apply it in the context of symbolic music generation using expert and learned features.
We train and release FIGARO, which combines two description functions in a model that achieves human-interpretable, fine-grained control on multi-track, multi-time-signature music while beating state-of-the-art controllable generation models both in terms of sample quality and style transfer.
% 2. both types of descriptions have trade-offs
The proposed description functions complement each other:
The expert description is a human-interpretable sequence that is easy to create and acts as a strong inductive bias to the model.
The learned description is able to mitigate non-injectivity present in the expert description and amend it with high-fidelity, detailed information.
In the case of conflicting descriptions, the model still produces plausible samples, thus preserving human-interpretability even in the presence of learned features.
We can combine both descriptions to achieve state-of-the-art symbolic music generation with human-interpretable, fine-grained control.\\
On a broader perspective, we hope that the proposed method is a step toward facilitating artists in their creative process as well as enabling amateurs to express themselves by lowering the barrier of entry to music creation and making the process faster and easier overall.

\vspace{-.1cm}
\section*{Ethics Statement}
\vspace{-.15cm}
Automatic music generation may raise ethical concerns similar to those of large language models.
The models may exhibit biases toward a certain style of music and may not represent the music of marginalized cultures accurately.
In our case, the majority of training samples are western music, which is not necessarily representative of music at large.
The model might also reproduce copyrighted material that is present in the training data and potentially generate samples that infringe on copyright law.
Additionally, powerful music model capable of competing with human composers have the potential to create a big impact on the music industry.
In this regard, we believe that our contribution is a step toward human-AI collaboration rather than competition.

\bibliographystyle{iclr2023_conference}
\bibliography{iclr2023_conference}

\newpage
\appendix

\section{Implementation Details}
\label{app:implementation_details}
\subsection{Description-to-Sequence Model}
We largely follow the original paper for both of our Transformer auto-encoders.
Unless specified differently, hyperparameters are left unchanged.
One exception is the reduced context size of 256 tokens for the sake of training time.
In the same spirit, we also reduce the number of encoder layers to 4 but leave the number of decoder layers at 6.
We use relative positional embeddings \citep{huang_improve_2020} as this has been shown to be beneficial for symbolic music generation \citep{huang_music_2018}.
In addition to the positional embeddings, we also add a learned bar embedding and a learned beat-position embedding.
This puts a limit on the maximum number of bars (512 in our case). With a $4/4$ time signature at 120 BPM this equates to ~17 minutes of music, which we deem an acceptable limitation.
In the case of using both descriptions, we simply add their token embeddings before passing it to the encoder.
In total, the description-to-sequence model has 44.6 M trainable parameters.

In the presence of the learned description $F_\mathrm{learned}$, the discrete latent codes from the VQ-VAE feature extractor are embedded using the frozen codebook from the VQ-VAE instead of re-learning new embeddings. This is done to ensure stable training and reduce the number of trainable parameters.

\subsection{VQ-VAE Model}
For our bar-level VQ-VAE model, we use the same Transformer auto-encoder backbone as for the description-to-sequence model.
The final layer of the encoder is pooled as proposed by \citet{devlin_bert_2019} and then fed through a vector-quantization bottleneck, quantizing it to 16 latent codes from a codebook of size 2048.
We use a modified version of the sliced vector quantization scheme proposed by \citet{kaiser_fast_2018} as our discretization bottleneck.
Specifically, we decompose the latent representation $\zz$ into 16 slices $\zz_1, \dots, \zz_{16}$ and discretize each of them to a shared codebook $\mathcal{C}$ with $|\mathcal{C}| = 2048$ using the $k$-means discretization technique from the original paper \citep{oord_neural_2018}.
The latent vector is provided to the decoder through cross-attention.
We use a linear layer before and after vector-quantization to project between model space and latent space.
The VQ-VAE model has 43.7 M trainable parameters in total.

The model is trained on bar-level reconstruction by minimizing the canonical $\beta$-VQ-VAE loss without the auxiliary codebook loss \citep{oord_representation_2019} given by
\begin{equation*}
    \mathcal{L}_{\mathrm{VQ-VAE}}(\phi) =
    \mathbb{E}_{\xx \sim \mathcal{X}}\left[
        -\log p_\phi(\xx | z_q(\xx)) + \beta\|z(\xx) - \mathrm{sg}(z_q(\xx)\|
    \right]
\end{equation*}
where $\mathrm{sg}(x)$ denotes the stop-gradient operator, $\phi$ are the model parameters and $\beta = 0.02$.
The codebook is updated using the EMA update step as proposed in the original paper.
We employ random restarts \citep{dhariwal_jukebox_2020} to ensure optimal codebook usage.
Bars with more tokens than the context size are truncated to fit.

\subsection{Training Details}
We train each model for 100k steps with a batch size of 512 sequences.
Models are optimized using the Adam optimizer \citep{kingma_adam_2017} with $\beta_1 = 0.9$, $\beta_2 = 0.999$, $\epsilon = 10^{-6}$ and $0.01$ weight decay.
We use the inverse-square-root learning rate schedule with initial constant warmup at $10^{-4}$ given by $10^{-4} /\max(1, \sqrt{n/N})$ where $N = 4000$ is the number of warmup steps.
For additional details and hyperparameters, we refer to the source code.

\section{REMI+ Input Representation}
\label{app:remi_plus}

We extend the original REMI representation \citep{huang_pop_2020} to make it suitable for general multi-track, multi-signature symbolic music sequences.
We make modifications that add time signature and instrument information, we determine a unique order of events and use quantization schemes that allow for accurate representation for a diverse set of music.
An example of a REMI+ sequence is given in Figure~\ref{fig:remi_example}.

\textbf{Bar Tokens.}
To provide the model with additional context, we include the index of the current bar in each bar token.
This, along with the bar embedding, should help the model retrieve the correct information from the description and help determine the end of a piece at generation time.

\textbf{Time signature.}
We add a time signature token at the beginning of each bar, indicating the time signature of the bar which is to follow. 
We adapt the convention that time signature changes may only happen at the beginning of a bar, which is commonly true in written music.

\textbf{Instruments.}
We add instrument information as an additional token before each note event, indicating which instrument will play the following note.

\textbf{Order of events.}
In theory, the order of notes within each bar can be arbitrary without compromising the validity of the sequence.
But to make the modelling task easier, we define a unique and deterministic order of events.
Specifically, we sort events by (Bar, Position, EventType, Instrument, Pitch) in ascending order (valid event types are \{Chord, Tempo, Note\}).
This order is unique since a given instrument can only ever play a single note with a given pitch at a given time\footnotemark.

\footnotetext{This is an assumption that could be violated in theory but does hold in practice.}

\textbf{Quantization.}
We largely follow \citet{huang_pop_2020} in quantization.
The most significant deviation is the use of 12 note onset positions per quarter note instead of 4 as proposed in the original work.
For example, there will be 48 unique note onset positions for the $4/4$ time signature and 36 note onset positions for the $3/4$ time signature.
This allows both triplet and sixteenth notes to be quantized accurately, which is important when considering a diverse set of music.

Instruments and note pitches are not quantized as they are categorical variables with 128 possible values by the MIDI specification.
Note velocity is quantized to 32 intervals in $[0, 128]$ and note duration is quantized to position intervals defined by the following mesh:
\begin{align*}
    \label{eq:note_duration_quant}
    \mathcal{M} =\; &\{1, \dots, 12\} \;\cup \\
    &\{12 + 3i \mid i \in (1, \dots, 4)\} \;\cup \\
    &\{12 + 4i \mid i \in (1, \dots, 3)\} \;\cup \\
    &\{24 + 6i \mid i \in (1, \dots, 4)\} \;\cup \\
    &\{48 + 12i \mid i \in (1, \dots, 12)\} \;\cup \\
    &\{192 + 24i \mid i \in (1, \dots, 24)\}
\end{align*}
This ensures single position accuracy up to quarter notes, then $16$th and triplet accuracy up to half notes and $8$th note accuracy up to a full note.
To limit vocabulary size, we switch to quarter note steps up to $8$ full notes and half note steps up to $16$ full notes after that.
Notes longer than $16$ full notes are truncated to this length.
Finally, tempo change events are discretized to 32 intervals in $[0, 240]$.

\begin{figure}[H]
    \centering
    \begin{tabular}{l}
    \scriptsize
\begin{lstlisting}
<bos>
Bar_1 TimeSignature_3/4
  Pos_0 Tempo_120
  Pos_0 Chord_C:min
  Pos_0 Instrument_Drums Pitch_36 Vel_90 Dur_0
  Pos_0 Instrument_Piano Pitch_64 Vel_85 Dur_4
  Pos_4 Instrument_Piano Pitch_66 Vel_85 Dur_4
Bar_2 TimeSignature_3/4
  Pos_0 Tempo_120
...
<eos>\end{lstlisting}
    \end{tabular}
    \caption{Example sequence represented in the REMI+ representation. At the beginning of each bar time signature, tempo and the current chord are noted, after which each note is represented through five subsequent tokens.}
    \label{fig:remi_example}
\end{figure}

\section{Expert Description Algorithm}
\label{description_details}

Pseudocode for generating the expert description is given in Algorithm~\ref{alg:expert_description}.
We quantize note density, mean pitch and velocity to 32 linearly spaced intervals in $[0, 12]$, $[0, 128]$ and $[0, 128]$ respectively.
Mean duration is quantized to 32 logarithmically spaced intervals in $[0, 128]$ positions (12 positions per quarter note).
Chords are extracted with an adapted version of the Viterbi algorithm also used by \citet{huang_pop_2020}.

\newcommand{\CALL}[2]{\textsc{#1}(#2)}
\begin{algorithm}
    \caption{ExpertDescription}
    \label{alg:expert_description}
    \begin{algorithmic}
        \STATE \textbf{input} musical sequence $\xx$
        \STATE \textbf{output} description $\mathbf{d}$
        \STATE $\mathbf{d} \gets ()$
        \STATE $b_1, \dots, b_n \gets $ \CALL{PartitionIntoBars}{$\xx$}
        \FOR{$b_i \in (b_1, \dots, b_n)$}
            \STATE $N \gets \{n \mid \text{$n$ is a note with onset in $b_i$}\}$
            \STATE $I \gets \{\texttt{inst} \mid \text{\texttt{inst} is being played during $b_i$}\}$
            \STATE $C \gets \{\texttt{chord} \mid \text{\texttt{chord} is being played during $b_i$}\}$
            \STATE $q \gets $ duration of $b_i$ in quarter notes
            \STATE $\texttt{ts} \gets$ time signature at beginning of $b_i$
            \STATE $\texttt{nd} \gets \frac{|N|}{q}$
            \STATE $\texttt{mp} \gets \frac{1}{|N|} \sum_{n \in N}$ \CALL{Pitch}{$n$}
            \STATE $\texttt{mv} \gets \frac{1}{|N|} \sum_{n \in N}$ \CALL{Velocity}{$n$}
            \STATE $\texttt{md} \gets \frac{1}{|N|} \sum_{n \in N}$ \CALL{Duration}{$n$}
            \STATE Quantize \texttt{nd}, \texttt{mp}, \texttt{mv} and \texttt{md}
            \STATE $d_i \gets (i, \texttt{ts}, \texttt{nd}, \texttt{mp}, \texttt{mv}, \texttt{md}) \parallel \mathrm{list}(I) \parallel \mathrm{list}(C)$
            \STATE $\mathbf{d} \gets \mathbf{d} \parallel d_i$
        \ENDFOR
        \STATE \textbf{return} $\mathbf{d}$
    \end{algorithmic}
\end{algorithm}

\section{Evaluation Metrics}
\label{app:evaluation_metrics}

\subsection{Macro Overlapping Area}
As used by previous work, the overlapping area (OA) metric does not consider the sequential order of the investigated feature, as feature histograms are computed over the entire sequence.
For example, the overlapping area of $\xx$ and $\mathrm{reverse}(\xx)$ would be maximal, even though the reversed sequence does not sound like the original sequence in general.

To alleviate this limitation, we adapt the OA metric to also consider temporal order.
Let $\xx$ and $\yy$ denote two musical sequences and let $b_i^{(\xx)}$ and $b_i^{(\yy)}$ denote the $i$-th bar of $\xx$ and $\yy$ respectively.
We compute the overlap in feature distributions for each bar by fitting a Gaussian distribution to the feature under examination (e.g. note pitch) and compute the overlapping area between the two distributions.
Let this overlap be given by $\mathrm{overlap}(b_i^{(\xx)}, b_i^{(\yy)})$.
Then the macro overlapping area (MOA) between $\xx$ and $\yy$ is given by
\begin{equation*}
    \mathrm{MOA}(\xx, \yy) = \frac{1}{N} \sum_{i=1}^N \mathrm{overlap}\left(b_i^{(\xx)}, b_i^{(\yy)}\right)
\end{equation*}

\subsection{Normalized Root-Mean-Square Error}
In order to normalize for different feature magnitudes between different samples, we compute the normalized RMSE (NRMSE) for bar-wise note density.
Let $x$ denote the ground truth, $\hat{x}$ denote the reconstruction and $N$ denote the length of the sequences.
Then the NRMSE is given by
\begin{align*}
    \mathrm{RMSE}(x, \hat{x}) &= \sqrt{\frac{1}{N}\sum_{i=1}^N (\hat{x}_i - x_i)^2} \\
    \mathrm{NRMSE}(x, \hat{x}) &= \frac{\mathrm{RMSE}(x, \hat{x})} {\mathrm{mean}(x)}
\end{align*}

\subsection{Cosine Similarity}
Let $\mathbf{v}^{(\xx)}_i$ and $\mathbf{v}^{(\yy)}_i$ denote the chroma vector \citep{fujishima_real-time_1999} or grooving vector \citep{dixon_towards_2004} for the $i$-th bar in $\xx$ and $\yy$ respectively.
We then average the cosine similarity over the entire sequence to get the chroma/grooving similarity:
\begin{equation*}
    \mathrm{sim}_\mathbf{v}(\xx, \yy) = \frac{1}{N} \sum_{i=1}^N \frac{\mathbf{v}_i^{(\xx)} \cdot \mathbf{v}_i^{(\yy)}} {\|\mathbf{v}_i^{(\xx)}\|\|\mathbf{v}_i^{(\yy)}\|}
\end{equation*}

\section{Benchmark Models}
\label{app:benchmark_models}
\subsection{\citet{huang_music_2018}}
We reimplement the model proposed by \citet{huang_music_2018} as an unconditional baseline.
We train the model on the REMI+ representation to allow for direct comparison to our method.
We use the improved relative attention from \citet{huang_improve_2020} to eliminate any possible advantage arising from different attention mechanisms.
We largely stick to the hyperparameters from our models, using 6 decoder layers with a hidden size of 512 and a filter size of 2048.
Training and optimization hyperparameters are also the same as for our models.

\subsection{\citet{choi_encoding_2020}}
We reimplement the model proposed by \citet{choi_encoding_2020} and train it on the REMI+ representation to allow for direct comparison to our method.
We again use the improved relative attention from \citet{huang_improve_2020} and largely stick to the hyperparameters from the original paper, using 6 encoder and 6 decoder layers.
Unlike the original work, we do not use data augmentation since the dataset is large enough and in order to allow for fair comparison between the models.
Due to GPU memory constraints we reduce the context size from 2048 to 1024 and use an accumulated batch size of 16, ensuring stable training.
Training and optimization hyperparameters are the same as for our models.

\subsection{\citet{wu_musemorphose_2021}}
We train MuseMorphose on the REMI+ representation to allow for direct comparison to our method.
Adapting the released implementation for our experiments, we reduce the context size from 1280 to 512 tokens due to GPU memory constraints but leave all other hyperparameters as they were proposed in the original paper.
The model is trained until convergence (approx. 125k steps).
We limit the training data to a subset of the entire dataset (20k samples) due to technical limitations.
This is still considerably more training data than what was used in the original paper (1k samples) and should not affect performance significantly compared to using the full dataset.

\section{Description Mixing}
\label{app:description_mixing}
\begin{table}
    \centering
    \begin{tabular}{|l|ccc|cccccc|}
        \hline
         & \multicolumn{3}{c|}{Accuracy} & \multicolumn{6}{c|}{Fidelity} \\
         & I\;$\uparrow$ & C\;$\uparrow$ & TS\;$\uparrow$ & ND\;$\downarrow$ & P\;$\uparrow$ & V\;$\uparrow$ & D\;$\uparrow$ & $s_c\uparrow$ & $s_g\uparrow$ \\
        \hline
        FIGARO & .960 & .593 & .997 & .238 & .827 & .735 & .748 & .790 & .853 \\
        \hline
        Expert Similarity & \textbf{.819} & \textbf{.351} & \textbf{.993} & \textbf{.453} & .625 & .354 & .473 & \textbf{.595} & .621 \\
        Learned Similarity & .288 & .057 & .753 & .696 & \textbf{.699} & \textbf{.670} & \textbf{.623} & .398 & \textbf{.741} \\
        \hline
        Mean & .554 & .204 & .873 & .575 & .662 & .512 & .548 & .497 & .681 \\
        Best case & .819 & .351 & .993 & .453 & .699 & .670 & .623 & .595 & .741 \\
        \hline
    \end{tabular}
    \vspace{0.3cm}
    \caption{Description mixing performance on similarity metrics. \enquote{FIGARO} refers to the performance on unaltered descriptions. Expert/learned similarity refers to similarity of generated samples to the sequence from which the expert/learned description was extracted from. Highlights indicate which of the two descriptions was more closely adhered to in terms of the respective metric.}
    \label{tab:description_mixing}
\end{table}
Quantitative evaluation for this task is not straight-forward as there are two different \enquote{ground truth} samples with potentially conflicting features and it is unclear, which one the model should adhere to for which features.
It is clear though that the model follows one of the descriptions more closely for some features, which points to some degree of separation of concerns between the two descriptions.
Exact numbers are provided in Table~\ref{tab:description_mixing}.

\section{Subjective Evaluation}
\label{app:subjective_evaluation}
\begin{table*}[t]
    \centering
    \begin{tabular}{llclr}
        \toprule
        Model & Opponent & Winrate & $p$-value & $N$ \\
        \midrule
        1. Ground truth & & $0.669$ & & $2418$ \\
        \midrule
        & FIGARO & $0.605$ & $< 10^{-6}$ & $595$ \\
        & \citet{huang_music_2018} & $0.663$ & $< 10^{-15}$ & $605$ \\
        & \citet{choi_encoding_2020} & $0.647$ & $< 10^{-12}$ & $586$ \\
        & \citet{wu_musemorphose_2021} & $0.756$ & $< 10^{-37}$ & $632$ \\
        \midrule
        2. FIGARO & & $0.581$ & & $2439$ \\
        \midrule
        & \citet{huang_music_2018} & $0.609$ & $< 10^{-7}$ & $635$ \\
        & \citet{choi_encoding_2020} & $0.624$ & $< 10^{-9}$ & $631$ \\
        & \citet{wu_musemorphose_2021} & $0.696$ & $< 10^{-20}$ & $578$ \\
        \midrule
        3. \citet{huang_music_2018} & & $0.463$ & & $2467$ \\
        \midrule
        & \citet{choi_encoding_2020} & $0.543$ & $0.0162$ & $613$ \\
        & \citet{wu_musemorphose_2021} & $0.581$ & $< 10^{-4}$ & $614$ \\
        \midrule
        4. \citet{choi_encoding_2020} & & $0.447$ & & $2490$ \\
        \midrule
        & \citet{wu_musemorphose_2021} & $0.589$ & $< 10^{-5}$ & $660$ \\
        \midrule
        5. \citet{wu_musemorphose_2021} & & $0.345$ & & $2484$ \\
        \bottomrule
    \end{tabular}
    \caption{Ranking of the different methods by winrate according to the user study. The Wilcoxon signed-rank test is used to calculate pairwise ranking $p$-values. \enquote{Ground truth} denotes sampling the test set.}
    \label{tab:wilcoxon_test}
\end{table*}

\begin{figure}
    \centering
    \begin{subfigure}{\textwidth}
        \centering
        \includegraphics[width=\textwidth]{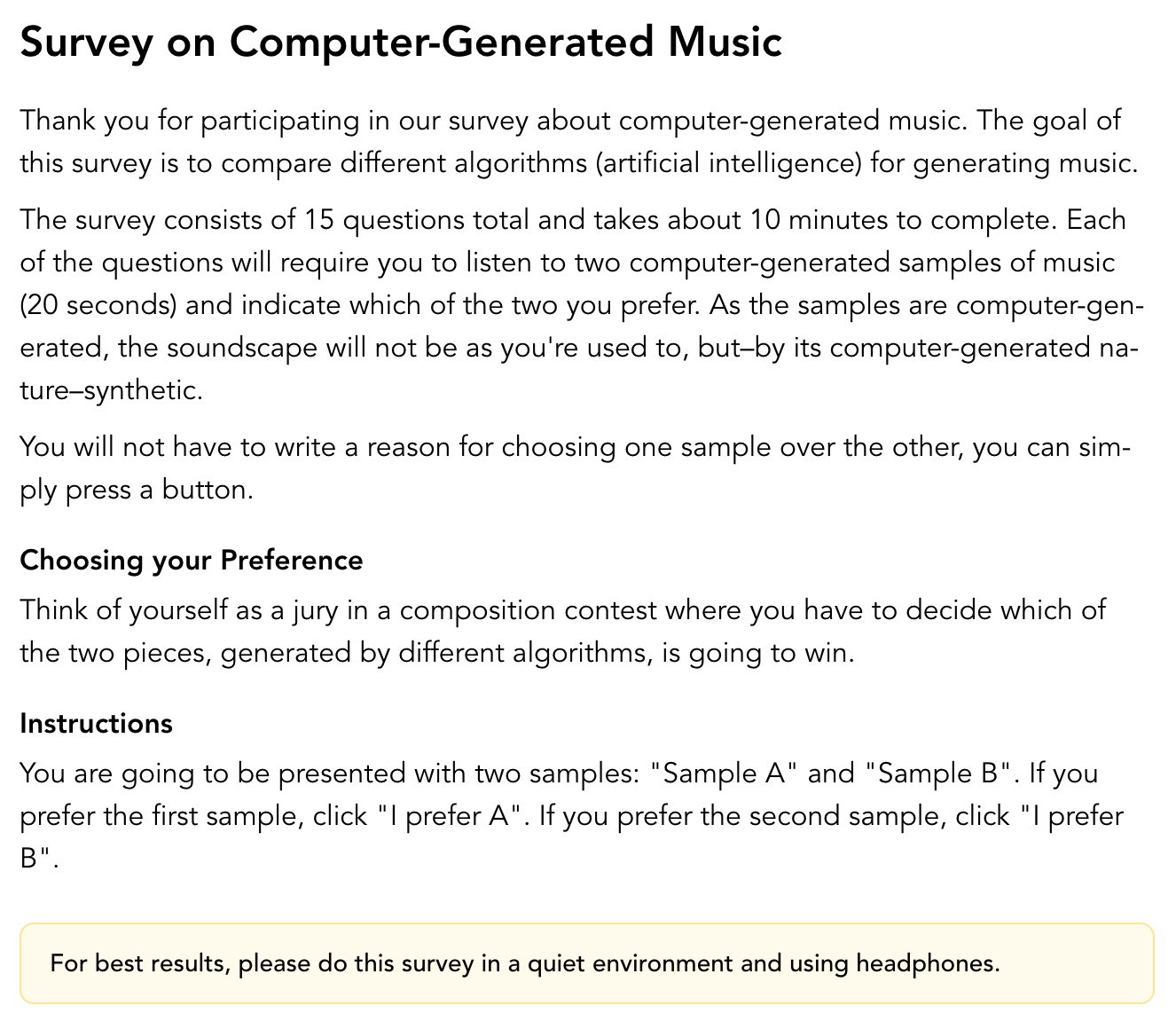}
        \caption{Instructions that were provided to the participants of the listening study.}
    \end{subfigure}
    \begin{subfigure}{\textwidth}
        \vspace{.5cm}
        \centering
        \includegraphics[width=\textwidth]{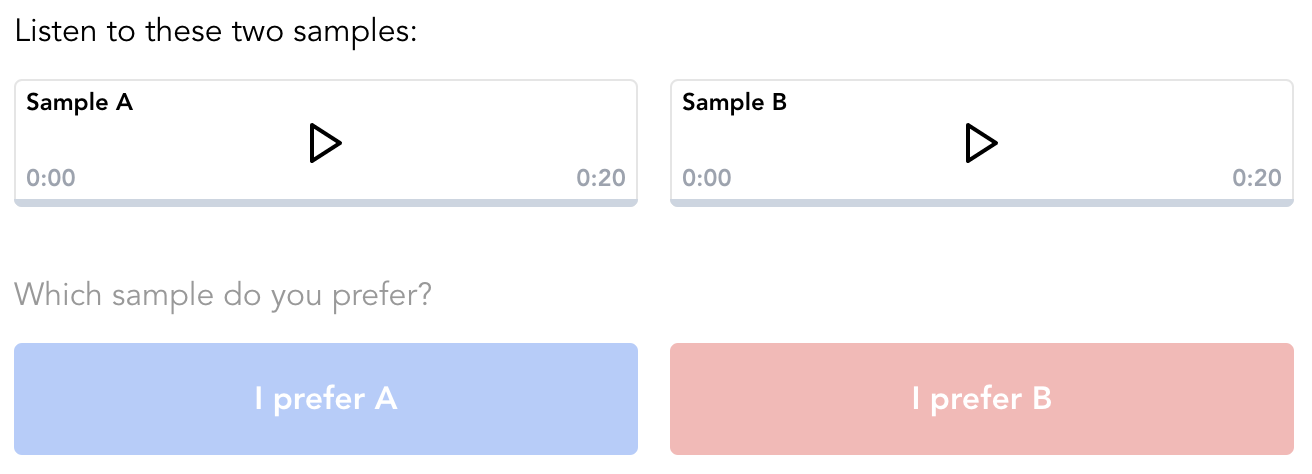}
        \caption{Screenshot of the survey questions that were provided to the study participants after reading the instructions.}
    \end{subfigure}
    \vspace{.3cm}
    \caption{Screenshots of the listening study. Participants were first presented with the instructions (a) before answering 15 questions of the type presented in (b).}
    \label{fig:screenshots}
\end{figure}

The listening study includes 7569 comparisons by 691 participants, each averaging 11 answers.
In each comparison, participants were presented with two different samples and were asked to indicate, which of the two they preferred.
For the following ranking test, samples were chosen uniformly at random from two different generative models, the pair of which again was chosen at random.
In this setup, we treat real samples as one possible model, for which we simply sample the test distribution.
For the other models, we generate samples as described for the style transfer experiment (Section~\ref{sec:conditional_generation}).
Screenshots from the study are presented in Figure~\ref{fig:screenshots}.

We apply the Wilcoxon signed-rank test to the results in order to establish a ranking of the different models.
FIGARO beats each baseline except ground truth with a win rate of more than $60\%$ and a $p$-value of less than $10^{-7}$.
Out of all methods, FIGARO also has the highest win rate against real samples with a win rate of $39.3\%$, which is a $6\%$ advantage over the next best method.
Rankings and corresponding $p$-values are reported in Table~\ref{tab:wilcoxon_test}.

\end{document}